\documentclass[11pt]{article}

\usepackage{amssymb}
\usepackage{amsmath}
\usepackage{cite}
\usepackage{epsf}

\newcommand{\sfrac}[2]{{\textstyle\frac{#1}{#2}}}

\renewcommand{\thefootnote}{\arabic{footnote}}
\newcommand\ZZ{\hbox{\zfont Z\kern-.4emZ}}
\font\zfont = cmss10 
\newcommand{\Tt}{{\tilde t}}
\newcommand{\Tr}{{\tilde r}}
\newcommand{\hA}{{\hat A}}
\newcommand{\hB}{{\hat B}}
\newcommand{\hC}{{\hat C}}
\newcommand{\hr}{{\hat r}}

\begin{document}

\begin{titlepage}
\begin{flushright}
LBNL--47190\\
UCB--PTH--00/42\\
{\tt hep-th/0012143} \\
\end{flushright}

\vskip.5cm
\begin{center}
{\huge{\bf Gravitational Lorentz Violations and}}
\vskip.3cm
{\huge{\bf  Adjustment of the Cosmological Constant  }}
\vskip.2cm
{\huge{\bf in Asymmetrically Warped Spacetimes}}
\vskip.2cm
\end{center}
\vskip0.2cm

\begin{center}
{\sc Csaba Cs\'aki}$^{a,}$\footnote{J. Robert Oppenheimer 
fellow.}, {\sc Joshua Erlich}$^{a}$ and
{\sc Christophe Grojean}$^{b,c}$ \\

\end{center}
\vskip 10pt

\begin{center}
$^{a}$ {\it Theory Division T--8,
Los Alamos National Laboratory, Los Alamos, NM 87545, USA} \\ 
\vspace*{0.1cm}
$^{b}$ {\it Department of Physics,
University of California, Berkeley, CA 94720, USA} \\ \vspace*{0.1cm}
$^{c}$ {\it Theoretical Physics Group,
Lawrence Berkeley National Laboratory, \\ Berkeley, CA 94720, USA} \\ 
\end{center}

\vglue 0.3truecm

\begin{abstract}
\vskip 3pt \noindent
We investigate spacetimes in which the speed of light along flat 4D sections
varies over the extra dimensions
due to different warp factors for the space and the time coordinates
(``asymmetrically warped'' spacetimes). The main 
property of such spaces is that while the induced metric is flat, implying
Lorentz invariant particle physics on a brane, bulk gravitational effects
will cause apparent violations of Lorentz invariance and of causality
from the brane observer's point of view. 
An important experimentally verifiable
consequence of this is that gravitational waves may travel with a 
speed different from the speed of light on the brane, and possibly 
even faster. We find the 
most general spacetimes of this sort, which are given by AdS--Schwarzschild or
AdS--Reissner--Nordstr\"om black holes, assuming the simplest
possible sources in the bulk. Due to the gravitational Lorentz violations
these models do not have an ordinary Lorentz invariant effective description,
and thus provide a possible way around Weinberg's no-go theorem for the
adjustment of the cosmological constant. Indeed we show that the cosmological
constant may relax in such theories by the adjustment of the mass and
the charge of the black hole. 
The black hole singularity in these solutions
can be protected by a horizon, but the existence of a horizon requires
some exotic energy 
densities on the brane. We investigate the cosmological
expansion of these models and speculate that it may provide an explanation
for the accelerating Universe, provided that the timescale for the
adjustment is shorter than the Hubble time. In this case the accelerating 
Universe would be a manifestation of gravitational Lorentz violations in 
extra dimensions.

\noindent

\end{abstract}

\end{titlepage}

\renewcommand{\thefootnote}{(\arabic{footnote})}

\section{Introduction}
\setcounter{equation}{0}
\setcounter{footnote}{0}

The physics of extra dimensions has recently attracted renewed interest,
mainly for the following three reasons:
\begin{itemize}
\item{The existence of extra dimensions could provide a stable large hierarchy
between the scale of particle physics (the TeV scale) and the scale of 
gravity (the Planck scale)~\cite{large,RS}.}

\item{One can obtain the 4D Einstein equations from a higher dimensional
setup without compactification or a need for a stabilization
mechanism~\cite{RS,GT,GKR},
and the expansion of the brane driven by matter follows the usual
4D Friedmann equations~\cite{BraneExp}.}

\item{One may hope that the cosmological constant problem could be partly
resolved due to extra dimensional
physics~\cite{RubSha,ADKS,KSS,SelfGen,FLLN,CEGH,4Form,CCextra}}.
\end{itemize}

Randall and Sundrum (RS) studied spacetimes with a 
single ``warped'' extra dimension, for which the metric is of the form
\begin{equation}
ds^2=e^{-A(y)}\, \eta_{\mu\nu}\,dx^\mu dx^\nu+dy^2.
\label{warped}
\end{equation}
They showed that such geometries can have interesting consequences 
for particle physics and gravity.  In these scenarios the standard model fields
are usually assumed to live at a particular point in the extra dimension,
called a 3-brane.  Metrics of the form (\ref{warped}) can reproduce 
4D Einstein gravity on the brane without compactification, and can produce
a large hierarchy between the scales of particle physics and gravity due to
the appearance of the warp factor. 

In this paper we study the most general backgrounds with 
one extra dimension, assuming 3D rotational invariance 
is still maintained\footnote{The breaking of 3D rotational invariance
in the bulk would transmit its breaking in gravitational interactions
on the brane in contradiction with the isotropy of the cosmic
microwave background.}
(and with the additional assumption 
that the only sources in the
bulk are a bulk cosmological constant and a $U(1)$ gauge field). 
The metric of such backgrounds can be generically written as,
\begin{equation}
ds^2=-a^2(r,t) \,dt^2 + b^2(r,t)\, d\vec{x}^2 + c^2(r,t)\, dr^2.
\end{equation}
However, when the sources are restricted to a bulk cosmological constant
and a bulk gauge field, a five dimensional version of Birkhoff's theorem holds,
and such a metric can always be transformed into the form,
\begin{equation}
ds^2=-h(r)dt^2+r^2 d\Sigma^2+h(r)^{-1}dr^2.
\label{metric}
\end{equation}
Here $d\Sigma$ is the unit metric of the 3D sections $r,t$=const., and 
$h(r)$ describes
a black hole spacetime.
Examining (\ref{metric}) one can see that
the novel property of such a general 
5D spacetime compared to conventional warped background (\ref{warped}) 
is that the warp factors for the space and time components of
the 4D sections $r$=const. are generically different. 
We will refer to metrics of this form as  ``asymmetrically warped".
The induced metric at 
the 
4D sections may still be flat, implying that (up to quantum 
gravitational corrections)
particle physics on the brane
will see a Lorentz invariant spacetime. However, since every 4D section
of the metric (\ref{metric}) will have a differently defined Lorentz
symmetry (one needs to rescale the time coordinate differently at 
different points along the extra dimension to maintain a speed of light 
$c=1$ along each section),
the spacetime
(\ref{metric}) globally violates 4D Lorentz invariance, leading to
apparent violations of Lorentz invariance from the brane observer's 
point of view due to bulk gravity effects. These Lorentz 
violations
however produce different effects from explicit Lorentz violations that are
sometimes introduced (see for example~\cite{ColemanGlashow}) in particle
physics. In fact, since at the classical level they only affect 
gravity, the most striking consequence of this setup would 
be that the speed of gravitational waves would be different from the
speed of light, which could cause apparent violations of causality
from the 4D brane observer's point of view.
This possibility
has already been pointed out by K\"albermann
and Halevi~\cite{Kalbermann}, Chung and Freese~\cite{CF}, 
Ishihara~\cite{Ishihara}
and Chung, Kolb and Riotto~\cite{CKR}.
Note also that the existence of different speeds of propagation for
gravitational and electromagnetic interactions has some common
features with four dimensional theories of gravity with
two light cones as proposed in~\cite{bimetric}.
In asymmetrically warped spacetimes the reason for the different speeds of 
propagation for graviton and 
photon is
that in the background (\ref{metric}) 
the speed of light along the brane is changing as one is moving along the 
extra dimension.
Thus this setup is analogous to a medium with a changing index of refraction.
If the speed of light away from the brane is increasing, then by
Fermat's principle the geodesic between two points on the
brane will bend into the bulk, and the gravitational
wave which is not forced to propagate on the brane will arrive faster 
than the light signal which is stuck to the brane. In fact, this
difference in the speed of electromagnetic and gravitational 
waves could be viewed as a generic prediction of extra dimensions,
which can be experimentally verified~\cite{CKR,bimetric}
once gravitational waves are observed by LIGO, VIRGO or LISA.

It has been widely recognized starting with the work of Rubakov and
Shaposhnikov~\cite{RubSha}, that extra dimensions provide a new approach to
the 
cosmological constant problem: with the presence of extra dimensions
one no longer needs to ensure the vanishing of the 4D vacuum energy
on the brane in order to obtain a static flat brane. The non-vanishing
brane tension can be balanced by the bulk curvature, exactly as it happens
for example in the Randall-Sundrum model. However, in the RS background
a tuning between the bulk cosmological constant and the brane tension
is required in order to achieve this balance. The second important 
consequence of the metric (\ref{metric}) is that it contains parameters
in addition to the ones contained in the solution discussed by RS, 
namely the mass, charge and location of the black hole with respect to the 
brane, which can
be thought of as integration constants for the most general solution
in the bulk. Thus one may hope that the finetuning required in the 
original RS setup in order to ensure the vanishing of the 4D effective 
cosmological constant could be eliminated. This would be similar in spirit
to the ``self-tuning'' models~\cite{ADKS,KSS}, where a bulk scalar is
coupled to the tension
of the visible brane, and allows the transfer of curvature into the bulk if the
brane tension is adjusted, so as to maintain flatness of the brane. In our 
case, the adjustment mechanism would imply
that once the brane tension goes through a phase transition, the 
black hole would adjust its mass, charge and location in order to
balance the new stress tensor on the brane. One important ingredient in 
the model of~\cite{ADKS,KSS} was that for a particular coupling
of the bulk scalar field to the brane tension the only maximally symmetric
solution to the equations of motion was the flat brane solution, thus
making it plausible that the time dependent solution may relax to the
flat brane. In the black hole backgrounds this feature is 
automatically present, without having to tune a coupling: for generic
equations of state we will show that the only maximally symmetric brane
sections correspond to the flat branes. 
For simplicity
we will assume that the spacetime is $\mathbb{Z}_2$ symmetric around the brane.

There are two reasons 
why this approach based on bulk black hole spacetimes 
may be preferable compared to the scenario in~\cite{ADKS,KSS}:
\begin{itemize}
\item{Due to the gravitational Lorentz violations explained above, there is
no ordinary Lorentz invariant
low-energy effective 4D description of this model, and thus 
Weinberg's no-go theorem~\cite{Weinberg} for 4D adjustment mechanisms
of the cosmological
constant can not be applied here.  This reasoning is similar to that used to
argue that models of quasilocalized gravity~\cite{quasi} may be
relevant to the cosmological constant problem.}
\item{One may hope that due to the appearance of black holes, the
unavoidable~\cite{CEGH} naked singularities appearing in~\cite{ADKS,KSS}
would be replaced by black hole singularities shielded by a horizon.}
\end{itemize}
We will find that it is in fact possible to protect the singularities by a
horizon, and the spacetime can be cut at the location of the horizon
without reintroducing finetuning into the theory (that is, without the 
need to regulate the spacetime with, for example, an additional brane as
in~\cite{FLLN}). However,
we find that such horizons are consistent with the presence of a 
$\mathbb{Z}_2$ 
symmetric brane only in the case of charged black holes,
and require the presence of some exotic (but not fine-tuned) energy density
on the brane.  Of course this exotic energy density is not generically
required for an asymmetrically warped background, only for an adjusting
solution with a horizon. 

In order to gain some insight on how the effective gravity theory would behave
from the brane observer's point of view, we investigate some of the basic
features of the cosmology of these models. We find that the Friedmann
equation does contain the ordinary expansion term, but there are additional
contributions due to the mass and charge of the black hole,
which was also pointed out in \cite{Visser,CoHo}. These terms 
could be useful in explaining the apparent acceleration
of the Universe\cite{accelerating},
however in order to obtain a viable cosmology and solve the cosmological 
constant
problem at the same time 
the adjustment timescale 
of the mass and charge of the black hole must be shorter than the Hubble 
time. 

The paper is organized as follows: in Section 2 we first show what the
generic form of the solutions in the bulk is, and then introduce the brane
by solving the junction conditions. We examine the fine-tuning of the 
parameters and the existence of a horizon. We close Section 2 by showing
that there are no other maximally symmetric solutions. Section 3 is 
devoted to the physical consequences of the asymmetrically warped 
spacetimes. We calculate the geodesics and the speed of gravitational waves,
then show how a graviton zero mode would look like in these theories, and
finally consider the cosmology of these models.

\section{Flat branes in black hole backgrounds}
\label{sec:self-tune}
\setcounter{equation}{0}
\setcounter{footnote}{0}

Our aim is to find general 4D Lorentz invariant brane (``flat brane") solutions
to Einstein's equations in five dimensions, and to investigate
the following issues: is it possible to find flat brane
solutions without naked singularities, without fine tuning the parameters of 
the theory, and what would
the general properties of such spaces be? In this section we will
first discuss the form of the solution in the bulk, and then impose the
condition for a static brane. We will see that the pure gravity theory
will always require a fine-tuning between the parameters of the theory
for a solution to exist. However, if one also includes a gauge field in the
bulk, the situation changes, and one can find flat solutions for a range
of parameters with no tuning. However, for these solutions one of the
following two limitations will apply: either there is no horizon but a
naked singularity, or if one does want to ensure the existence of a 
horizon one has to assume the presence of some exotic matter on the 
brane.

\subsection{The solution in the bulk}

We model the 5D bosonic sector of the theory by a graviton and a cosmological 
constant
$\Lambda_{bk}$ as in the Randall--Sundrum model, and in addition
a $U(1)$ gauge field propagating in the bulk --- we will assume that the 4D
matter
living on the brane is neutral under this gauge symmetry. So the action 
describing
the dynamics of the system is given by,\footnote{
Our conventions correspond to a mostly positive Lorentzian signature
$(-+\ldots +)$ and the definition of the curvature in terms of the metric is
such that a Euclidean sphere has  positive curvature. Bulk indices will be 
denoted
by Latin indices ($M,N\ldots$) and brane indices by Greek indices
($\mu,\nu\ldots$).}
\begin{equation}
	\label{eq:action}
{\cal S} =
\int d^5x\,\sqrt{|g_5|} \,\left(  \sfrac{1}{2\kappa_5^2} R - \sfrac{1}{4} F_{MN}F^{MN} - \Lambda_{bk} \right)
+
\int d^4x\,\sqrt{|g_4|}\, {\cal L}_{\rm mat.} (g_{\mu\nu},\psi^{(m)})
\end{equation}
where $g_5$ and $g_4$ are, respectively, the determinants the 5D metric
$g_{MN}$ and the 4D metric induced on the brane, $g_{\mu\nu}$;
$\psi^{(m)}$ denote some matter fields localized on the brane and
${\cal L}_{\rm mat.}$ describes their interactions, and we will assume
that this matter can be described as a perfect fluid of energy density
$\rho$ and pressure $p$;
$\kappa_5^2$ defines the 5D Planck scale.  Note that the brane Lagrangian
${\cal L}_{\rm mat}$ includes the brane tension $\rho=-p$. 

We will assume that the solution is homogeneous and isotropic in the three
spatial directions of the brane. The general ansatz is,
\begin{equation}
\label{generalmetric}
ds^2= -n^2(\Tt,\Tr) \,d\Tt^2
+ a^2(\Tt,\Tr) d\Sigma_k^2
+ b^2(\Tt,\Tr) \,d\Tr^2,
\end{equation}
where 
$d\Sigma_k^2= d\sigma^2/ (1-k L^{-2} \sigma^2) +\sigma^2 d\Omega_2^2$
is the metric of the spatial 3-sections, with curvature parameter $k$,
$L$ being a parameter with dimension of length that will be
set to the length scale given by the cosmological constant in the bulk.
The brane is located at $\Tr=0$.
The 3D sections $\tilde{r},\tilde{t}$=const.
are either
a plane ($k=0$), a unit sphere ($k=1$) or a unit hyperboloid ($k=-1$). The brane
will be expanding or contracting 
as long as the scale factor $a(\Tt,0)$ explicitly depends
on time. Some of the features of metrics of the above form 
(\ref{generalmetric})
have also been independently 
investigated in Refs.~\cite{Kalbermann,CF,CKR},

In the case when there is no gauge field in the bulk, Kraus~\cite{Kraus}
and later Bowcock {\it et al.}~\cite{Bowcock}
have shown that a generalization of Birkhoff's theorem~\cite{Bkff}
holds in the bulk, which implies that there exists a system of coordinates 
where the
5D metric is static while, in general, the brane is moving and expanding.
In this system of coordinates, the most general metric is,
\begin{equation}
	\label{eq:metric}
ds^2=
-h(r)\,dt^2
+l^{-2} r^2\, d\Sigma_k^2
+h(r)^{-1}\,dr^2,
\end{equation}
where
\begin{equation}
	\label{eq:AdSSch}
h(r)=k +\frac{r^2}{l^2}-\frac{\mu}{r^2};
\ \ \ l^{-2}=-\sfrac{1}{6} \kappa_5^2 \Lambda_{bk}.
\end{equation}
The new coordinates $r$ and $t$ are functions of the brane-based coordinates
$\Tr$ and $\Tt$ and thus the location of the brane is parametrized by
$r=R(t)$ and is solution of a differential equation we will give in the next section.
The bulk geometry describes an AdS--Schwarzschild hole (a ``black wall'' for
$k=0$)
 located at $r=0$ and spreading
in the three other spatial directions. The parameter $\mu$ is interpreted as the
mass  (in units where the five dimensional Planck scale is equal to one)
of the black hole.
When $\mu=0$, the metric simply describes 5D Anti-de Sitter
space and for a non-vanishing positive $\mu$ the $r=0$ singularity is hidden
behind a horizon $r=r_h>0$ where the metric becomes degenerate $h(r_h)=0$.

Birkhoff's theorem can also be generalized when the gauge field is turned on.
This way we obtain the AdS--Reissner--Nordstr\"om metric 
(see also~\cite{Visser}), which has the same
form as (\ref{eq:metric}), except that now the function $h$ also depends
on the charge $Q$ of the black hole:
\begin{equation}
	\label{eq:AdSRN}
h(r)=k + \frac{r^2}{l^2} - \frac{\mu}{r^2} + \frac{Q^2}{r^4}
\end{equation}
The non-vanishing component of the bulk field strength tensor $F$ of the
gauge field satisfies the Bianchi identities and is given by,
\begin{equation}
	\label{eq:gauge}
F_{tr}=\frac{\sqrt{6} Q}{\kappa_5 r^3}.
\end{equation}
Later we will mainly be interested in the
case $k=0$ because in that case the (3+1)-dimensional sections
$r$=constant are flat.
If the charge of a Reissner--Nordstr\"om black hole is too large,
there is no horizon. In our solutions with $k=0$, the existence of
an inner and an outer horizon gives the following upper bound
on the charge
\begin{equation}
	\label{eq:RNhorizon}
Q^4<\sfrac{4}{27}\mu^3 l^2.
\end{equation}
This follows from the fact that the position of a horizon corresponds
to the square root of a positive root of the cubic function
$f_3(x)=x^3/l^2-\mu x+Q^2$. As long as its discriminant is negative,
$f_3$ will have three roots and, as we will argue in Section \ref{sec:AdS-RN},
two of them are positive and thus they are associated to two horizons. 

\subsection{Matching at the brane}

We will assume that a codimension one brane separates two 
regions of the above discussed 5D black hole spacetimes.  That is, just like
in the original RS scenario, we are gluing two slices of the metric together
at the position of the brane. In order for this to be possible, one has
to satisfy the Israel junction conditions (also known as ``the jump equations'') at the brane.
We will restrict ourselves to solutions with a $\mathbb{Z}_2$ symmetry
between the two sides of the brane which means that we study the expansion of
a brane located at a fixed point of a $\mathbb{Z}_2$ orbifold.
A simple way of defining a $\mathbb{Z}_2$ symmetric spacetime by gluing
patches together has been explained in Ref.~\cite{CG}. The
idea is that for a metric of the form
\begin{equation}
	\label{eq:ABCmetric}
ds^2= -A^2(r) dt^2+B^2(r) d\Sigma_{\kappa}^2+C^2(r) dr^2
\end{equation}
one can find another solution using the fact that one still has
reparametrization invariance in the coordinate $r$, and thus
$\hA(\hr)= A (f(\hr)),
\hB(\hr)= B (f(\hr)),
\hC(\hr)= \pm C (f(\hr)) f'(\hr)$
still solves the Einstein equations.
Thus one can use this invariance to generate new solutions by gluing
a solution to its image under the reparametrization. A particularly simple
way of doing this is by picking $f(r)=r_0^2/r$, and the
identification
\begin{eqnarray}
& {\rm for} \ \  r \leq r_0 \ & A(r)=A_0(r), \ B(r)=B_0(r), \ C(r)=C_0(r);
\nonumber \\
& {\rm for} \ \  r \geq r_0 \ & A(r)=A_0(r_0^2/r), \ B(r)=B_0(r_0^2/r),
\ C(r)=C_0(r_0^2/r) \frac{r_0^2}{r^2},
\end{eqnarray}
which will
automatically ensure the continuity of the metric functions, and give an
$r \leftrightarrow r_0^2/r$ $\mathbb{Z}_2$ symmetry of the metric. This is
analogous to the $y \leftrightarrow -y$  $\mathbb{Z}_2$ symmetry in the Randall-Sundrum
model, and in fact coincides with it upon doing the coordinate transformation
$r=l e^{-y/l}$. From now on we will always assume the existence of such
a $\mathbb{Z}_2$ symmetry in order to simplify the equations and we will apply
the previous method on the metric defined by (\ref{eq:AdSRN}) between the black hole and
the brane.

In a general case, the brane will not be static, but instead it will be
moving in the bulk spacetime, expressing the fact that the observed
brane Universe is expanding (or shrinking). The junction conditions
for this case have been worked out in detail in Refs.~\cite{Kraus,Ida,STW,Bowcock}.
It is assumed that the brane follows a trajectory $R(\tau)$ in the
above bulk spacetime. The proper time $\tau$ as observed on the brane
is defined by the equation
\begin{equation}
\dot{t}^2\, h(R(\tau ))-\dot{R}^2(\tau) \, h^{-1} (R(\tau ))=1,
\end{equation}
which ensures that the induced metric on the brane will be of the
Friedmann--Robertson--Walker form
$ds_{ind}^2=-d \tau^2+R(\tau )^2 d \Sigma_k^2$.
Keeping the interior (the region next to $r=0$) of the BH space-time 
to prevent the volume of the extra dimension to diverge,\footnote{If one 
wants to
keep the exterior (the region next to $r=+\infty$) of the BH space-time, both 
signs
of the jump equations (\ref{eq:generaljump1})--(\ref{eq:generaljump2}) have 
to be flipped.
We will discuss this issue in more detail at the end of Section 
\ref{sec:AdS-RN}}
one then obtains the junction conditions~\cite{Bowcock}:\footnote{Similar
differential equations have been obtained in~\cite{KK} in a different
context where the brane just probes the embedding space-time and moves
along a geodesic. The following equations ensure the consistency
of Einstein equations, and are the Israel junction conditions for this 
setup.}
\begin{eqnarray}
	\label{eq:generaljump1}
&&
\rho=
\frac{6}{\kappa_5^2 R} \sqrt{h(R) +\dot{R}^2},
\\
	\label{eq:generaljump2}
&&
\sfrac{2}{3}\rho+p =
-\frac{2\ddot{R}+ h'(R)}{\kappa_5^2 \sqrt{h(R)+\dot{R}^2}},
\end{eqnarray}
where $\rho$ is the energy density of the brane, while $p$ is its pressure.
That is, the energy momentum tensor on the brane is given by
$T^A_B={\rm diag} (-\rho, p,p,p,0) \delta (\sqrt{g_{rr}}(r-R(\tau)))$.
As long as the brane is moving, {\it i.e.}, $\dot{R} \not =0$, one combination
of these jump equations is just equivalent
to the energy conservation equation on the brane,
\begin{equation}
\label{jump1}
\dot{\rho} + 3 (\rho+p)\frac{\dot{R}}{R}=0,
\end{equation}
and the remaining equation simply reads,
\begin{equation}
\label{jump2}
\frac{\dot{R}^2}{R^2}
= \sfrac{1}{36} \kappa_5^4\,\rho^2
-\frac{h(R)}{R^2}.
\end{equation}
Note that in the case we will be mostly interested in of a static brane
(with also $\rho$ constant), the conservation equation is trivially satisfied
and in this case the two jump equations are really independent
and the equations simplify.
One way of obtaining these equations would be to simply set
$\dot{R}=\ddot{R}=0$ into Eqs. (\ref{eq:generaljump1})--(\ref{eq:generaljump2}).
Alternatively, one can use
the simple
jump conditions derived by Bin\'etruy {\it et al.}~\cite{BDL}
for a general metric of the form
\begin{equation}
ds^2= -n^2(t,r)\, dt^2 +a^2(t,r)\, d\Sigma_k^2 + b^2(t,r)\, dr^2,
\end{equation}
where in our case $n=\sqrt{h(r)}, a=r/l$
and $b=1/\sqrt{h(r)}$.
The conditions for a static brane at $r=r_0$ are then simply~\cite{BDL}:
\begin{equation}
\frac{[a']}{a|b|}=-\frac{\kappa_5^2}{3} \rho, \ \ \
\frac{[n']}{n|b|}=\frac{\kappa_5^2}{3} (2 \rho +3 p),
\end{equation}
where the above functions should be evaluated at the location of the brane
and $[f']$ stands for the jump of the derivative of the function $f$
around the brane: $[f']=\lim_{\epsilon \to 0} (f'(r_0+\epsilon) - 
f'(r_0-\epsilon))$. For the above explained $\mathbb{Z}_2$ symmetric construction
$[a']=-2/l$, $[n']=-h'/\sqrt{h}$, where we have assumed that for
$r<r_0$ we are using the slice of the black hole metric that includes the
singularity. Thus the jump conditions that 
a static brane has to satisfy are given by
\begin{equation}
	\label{eq:hjump}
6 \sqrt{h(r_0)} =\kappa_5^2 \rho r_0 \ \ \mbox{and } \ \
18 h'(r_0) = -\kappa_5^4 (2+3\omega)\rho^2 r_0
\end{equation}
where we have defined $\omega=p/\rho$. Of course these equations agree with 
the more general
jump equations (\ref{eq:generaljump1})--(\ref{eq:generaljump2}).
Note that the energy density on the brane, $\rho$, has to be positive
to cut away the infinity.

\subsection{The AdS--Schwarzschild black hole solutions}
\label{sec:AdS-Sch}

Let us now apply the jump equations for a static brane obtained above to the simple
AdS--Schwarzschild black hole case (no vector field in the bulk).
Our motivation is to find a vacuum solution with
4D Lorentz invariance on the brane and thus we will study static brane 
solutions~\footnote{
There is another way to obtain a flat brane solution since a Minkowski
metric can be written as an open FRW space ($k=-1$) with a scale factor linear
in the cosmic time. However, as it will become evident in our general
analysis in Section \ref{sec:MaxSym}, such a solution will not satisfy the
jump equations except in some very special cases.}
with a vanishing induced curvature on the three dimensional spatial 
sections ($k=0$).
We will postpone
consideration of other metrics with maximally symmetric 4D sections
to Section \ref{sec:MaxSym}.
In an AdS--Schwarzschild black hole space-time,
the jump equations (\ref{eq:hjump}) are: 
\begin{equation}
	\label{bhjump}
36 \left( \frac{r_0^2}{l^2}-\frac{\mu}{r_0^2} \right)
=
\kappa_5^4\rho^2r_0^2,
\ \
36 \left( \frac{r_0^2}{l^2}+\frac{\mu}{r_0^2} \right)
=
-\kappa_5^4 (2+3\omega) \rho^2 r_0^2 .
\end{equation}
Combining these two equations we obtain an expression for the black hole
mass $\mu$:
\begin{equation}
	\label{mueq}
\mu=-\sfrac{1}{24} \kappa_5^4 (1+\omega)\rho^2 r_0^4.
\end{equation}
The black hole singularity is shielded by a horizon if
its mass, $\mu$, is positive;
otherwise there is a naked singularity in the metric at $r=0$. This
condition then translates into $\omega <-1$. Note that this condition
violates the positive energy condition on the brane. However, it is not
immediately obvious, whether such an exotic matter would necessarily lead
to an instability. $\omega < -1$ could for example be obtained if in addition 
to a positive brane tension there is a small negative matter energy density
(which is not accompanied by pressure, that is $\omega=0$ for this
component).
Just as a negative tension brane at an orientifold type fixed point is
stable, this violation of the null energy condition at the $\mathbb{Z}_2$
symmetric brane might not be harmful.
This issue clearly calls for further investigations. 
As a corollary of (\ref{mueq}),
one can also see
that with an ordinary brane tension ($\omega$=$-1$), one can
not obtain a horizon\footnote{A static brane in a 
Schwarzschild black hole background with a horizon was found in~\cite{GJS}.
However, this brane is not 4D Lorentz invariant but of positive spatial
curvature ($k=+1$).}.
Moreover, even for the equation of state
$\omega <-1$, the solution is fine-tuned.
The reason is that from the jump equations (\ref{bhjump})
$\mu$ can be expressed in terms of the AdS length $l$ as
\begin{equation}
\frac{\mu}{r_0^4}= \frac{1}{l^2}-\frac{\kappa_5^4 \rho^2}{36},
\end{equation}  
which when substituted back into (\ref{mueq}) gives the relation\footnote{
Of course this relation makes sense only if $\omega<-1/3$. For other
equations of state, it is impossible for the brane to remain static
in an AdS--Schwarzschild black hole space-time.}
\begin{equation}
\rho =\sqrt{\frac{-72}{1+3\omega}}\, \frac{1}{l \kappa_5^2},
\end{equation}
which is a fine-tuning for the size of the energy density on the brane,
and is the analog of the Randall-Sundrum finetuning condition 
for the brane tension of the positive tension brane. Thus in the
presence of the $\mathbb{Z}_2$ symmetry the case with a simple
AdS--Schwarzschild black hole requires fine-tuning, and is not adjusting
its vacuum energy to zero.

\subsection{Self-tuning AdS--Reissner--Nordstr\"om black holes}
\label{sec:AdS-RN}

We have seen above that the simple AdS--Schwarzschild black hole solution requires fine tuning.
We will now show that the situation is different for the
AdS--Reissner--Nordstr\"om black hole. First we note that there are no
additional jump conditions due to the presence of the bulk gauge 
field, as long as the SM matter living on the brane is not charged 
under this gauge symmetry. The reason is that the Maxwell equations in
the bulk are first order differential equations
in terms of the
field strength $F_{AB}$.
Thus there are no
second derivatives appearing in the equation, and as long as there are 
no delta function sources for this gauge field on the brane (which is ensured
with the above assumption of SM not transforming under the bulk U(1) gauge symmetry),
there will be no additional jump conditions. One may wonder how this fits
with the fact that the only non-vanishing component of the field-strength
tensor is $F_{tr}$, which from the point of view of the underlying 
vector field $A_M$ is obtained by taking a derivative with respect to $r$,
and naively seems to be odd under the $\mathbb{Z}_2$ symmetry. 
The point is that,
to have an action $\mathbb{Z}_2$ invariant,
one has to choose the different components of $A_M$ to have different
properties under the $\mathbb{Z}_2$. However, the $\mathbb{Z}_2$ parity
of $A_r$ is a matter of choice. The Lagrangian (\ref{eq:action}) will be
invariant under the $\mathbb{Z}_2$ symmetry with both positive or negative
parity assignements for $A_r$. If we want to avoid the appearance of an
extra jump condition for the gauge fields one should choose
$A_r$ to be even and $A_{t,x}$ to be odd under the $\mathbb{Z}_2$ parity.
Then $F_{tr}$ is even, and we recover the above 
conclusion that there is no additional jump condition for the 
gauge field. In fact, such vector fields are indeed present in 
supergravity theories~\cite{LutySundrum}. However, with such a parity
assignement there is an additional Chern-Simons term allowed in the 
action. The AdS RN solution remains a solution in the presence of this
extra term in the Lagrangian, but it will likely not be the most 
general solution anymore. It would be worthwhile to study the 
most general solutions in the presence of the additional Chern-Simons 
term that we are setting to zero in this paper.

\begin{figure}[!tb]
\centerline{\epsfxsize=15cm\epsfbox{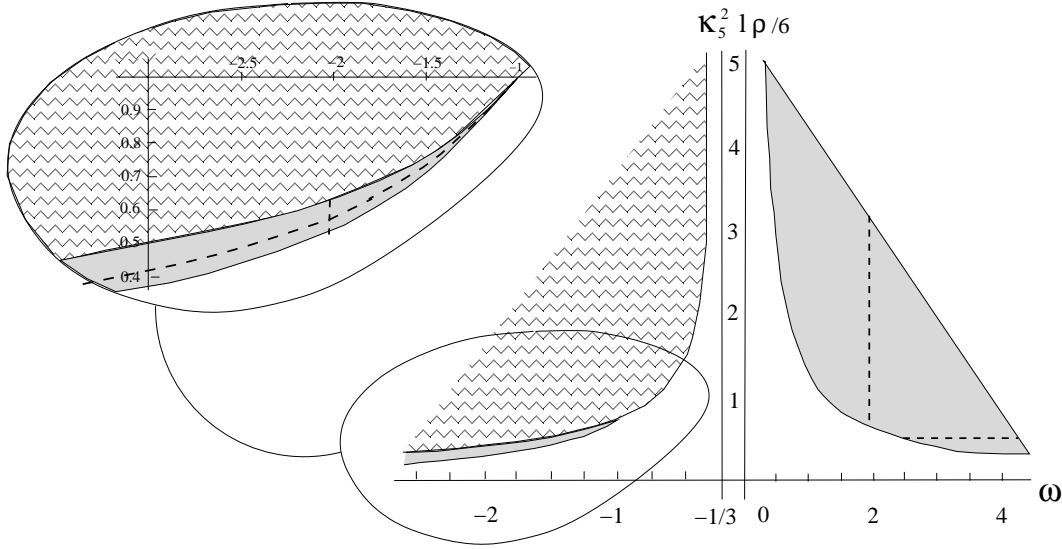}}
\caption[]{Phase diagram of a brane sitting in a bulk black hole background.
In the upper left zig-zagged region, the brane will expand. In the other parts
of the diagram, a brane can remain static and its induced metric is 4D Lorentz invariant.
In the other filled regions, the black hole singularity is hidden by two 
horizons; however
for $\omega>0$ these horizons are developed in the part of the BH space-time
cut by the $\mathbb{Z}_2$ orbifold. Fig.~\ref{fig:Horizons} shows the displacement
of the horizons when moving along the dashed lines varying the energy density or the
equation of state on the brane. The region $\omega<-1$ has been magnified.
}
\label{fig:Phase}
\end{figure}

The remaining part of this section is devoted to the embedding of a flat
brane in a charged black hole background. Not only can such 
a solution be found without finetuning of any parameter of the action, but
in some regions of the plane $(\omega,\rho)$ describing the brane, the
singularity of the black hole will be protected by two horizons. We have
summarized the results in Fig.~\ref{fig:Phase}.

The jump equations for a static brane embedded in a charged black hole
space-time are now
\begin{equation}
	\label{RNjump}
36 \left( \frac{r_0^2}{l^2}-\frac{\mu}{r_0^2}+\frac{Q^2}{r_0^4} \right)
=
\kappa_5^4\rho^2r_0^2,
\ \
36 \left( \frac{r_0^2}{l^2}+\frac{\mu}{r_0^2}-2\frac{Q^2}{r_0^4} \right)
=
-\kappa_5^4 (2+3\omega) \rho^2 r_0^2.
\end{equation}
The existence of the charge as a second constant of integration allows us 
to evade the previous
fine-tuning because the two jump equations simply fix the mass and the charge
of the black hole (in the case of a flat brane, {\it i.e.}, $k=0$) in terms of
brane parameters and $r_0$:
\begin{equation}
	\label{eq:RNmuQ}
\mu = 3\left( l^{-2} +\sfrac{1}{36} \kappa_5^4 \omega \rho^2 \right) r_0^4,
\ \ \
Q^2 = 2\left( l^{-2} +\sfrac{1}{72} \kappa_5^4 (1+3\omega)\rho^2 \right) r_0^6.
\end{equation}
When the parameter $\omega$ is too small, $\omega<-1/3$, the positivity
of $Q^2$ requires
\begin{equation}
	\label{eq:rho0}
\rho  \leq \rho_0=\sqrt{\frac{-72}{1+3\omega}}\, \frac{1}{l \kappa_5^2}.
\end{equation}

The black hole singularity at $r=0$ will be hidden behind a horizon
provided the inequality (\ref{eq:RNhorizon}) is fulfilled, which,
in terms of the equation of state and the energy density on the brane,
reads\footnote{Note that this inequality automatically implies
that the mass of the black hole is positive.}
\begin{equation}
	\label{eq:delta3}
l^{-2} \left( l^{-2} -\sfrac{1}{36}\kappa_5^4\rho^2 + \sfrac{1}{24}\kappa_5^4 (1+\omega)\rho^2  \right)^2
<
\left( l^{-2} -\sfrac{1}{36}\kappa_5^4\rho^2 + \sfrac{1}{36}\kappa_5^4 (1+\omega)\rho^2  \right)^3 .
\end{equation}
We immediately see that with an ordinary brane tension equation of state, 
$\omega$=$-1$,
this equation can never be satisfied.
However, as long as $\omega\not =-1$, there may exist in general an interval
for the energy density where the singularity is protected behind inner and 
outer horizons.
The constraint (\ref{eq:delta3}) for the existence of horizons can be written as a quadratic
inequality for the energy density,
\begin{equation}
4
+\sfrac{1}{36}l^{2}\kappa_5^4(1+6\omega-3\omega^2)\rho^2
-\sfrac{1}{324}l^{4}\kappa_5^8\omega^3\rho^4
<0,
\end{equation}
whose discriminant is $\Delta_2=(1+\omega)^3(1+9\omega)\kappa_5^8 l^{4}/1296$.
When the discriminant $\Delta_2$ is positive, the quadratic
equation will have two real roots. It can be checked that both
roots are positive when $\omega\leq -1$, both are negative when
$-1/9 \leq \omega \leq 0$ and only one is positive when $\omega\geq 0$. 
Since only
positive roots can correspond to a physical value of $\rho^2$, we conclude that
if $\omega$ is positive, the inequality is satisfied
for large enough value of the energy density
while in the negative range, $\omega$ has to be less than $-1$ and
the interval for the energy density where two horizons are developed
is included in
\begin{equation}\label{eq:rho_pm}
\rho_-<\rho<\rho_+
\ \ \mbox{with }
\ \
\rho_\pm=
\sfrac{6 }{l\kappa_5^2}\sqrt{\sfrac{1}{8\omega^3}(1+6\omega-3\omega^2 \mp \sqrt{(1+\omega)^3(1+9\omega)})}.
\end{equation}
However this interval has to be reduced further since we need to require
that $Q^2$ computed from (\ref{eq:RNmuQ}) is positive, {\it i.e.},
$\rho<\rho_0=\sqrt{-72/(1+3\omega)}/(l\kappa_5^2)$ which only partially
overlaps with the allowed interval.\footnote{ In order to have two horizons, we also
need the roots of the cubic equation associated to $h$ to be positive.
When the discriminant is negative, {\it i.e.}, the inequality (\ref{eq:delta3}) is satisfied,
the cubic equation has three real roots $x_1,x_2$ and $x_3$ with the properties:
$x_1+x_2+x_3=0$, $x_1x_2+x_2x_3+x_3x_1=-l^2 \mu/r_0^4$ and $x_1x_2x_3=-l^2 Q^2/r_0^6$.
Since furthermore the sign of the cubic function at the origin is related
to the sign of $Q^2$, we conclude that in the interval
(\ref{eq:iff}), the black hole singularity will be shielded by {\it two} horizons.
For $\omega<0$ and $\rho>\rho_0$, only one horizon would be present but
these solutions are non-physical since $Q^2<0$.}
In summary, the singularity at $r=0$ will be protected by horizons
{\it iff}
\begin{equation}
	\label{eq:iff}
(\omega>0 \ \ \mbox{and } \ \ \rho>\rho_-)
\ \ \ \mbox{or }
\ \ \
(\omega<-1 \ \ \mbox{and } \ \ \rho_-<\rho< \rho_0).
\end{equation}
We still have to impose that these horizons sit between
the singularity and the brane where the space is cut. This last condition
translates in an upper bound for the parameter $\omega$ defining the
equation of state. This condition is obtained
by studying the position of the brane with respect to the positions
of the horizons that correspond to the positive root of the cubic function
$f(x)=x^3/l^2-\mu x + Q^2$. It is worth noticing that the sign of $f$ is changing
when crossing a horizon but, from the first jump equation (\ref{eq:hjump}), we already
know that $f$ is positive at the brane. Moreover, $f'(0)=-\mu<0$ and the sign of $f'$
flips only between the two horizons. Thus it is enough to require that
$f'$ is positive at the brane, {\it i.e.},
$3r_0^4/l^2-\mu>0$ which from the expression of $\mu$ simply reads
$\omega<0$.

In conclusion, it is possible without any fine-tuning to embed
a static brane in a charged black hole bulk. Moreover the singularity
of this background will be hidden behind horizons if
\begin{equation}\label{eq:range}
\omega<-1 \ \ \mbox{and } \ \
\sfrac{6 }{l\kappa_5^2}\sqrt{\sfrac{1}{8\omega^3}(1+6\omega-3\omega^2 +
\sqrt{(1+\omega)^3(1+9\omega)})}<\rho<\sqrt{\frac{-72}{1+
3\omega}}\frac{1}{l\kappa_5^2} .
\end{equation}
Pictorially, the {\it phase diagram} of a brane in bulk black hole background
is summarized in Fig.~{\ref{fig:Phase}}
Clearly, the same comments apply here for the $\omega < -1$ matter as for 
the AdS Schwarzschild case. We should note, that it is not clear whether 
or not the existence of such exotic matter is a generic requirement 
for a more complicated
asymmetrically warped background 
(which could be obtained by introducing more
fields into the bulk) to be self-adjusting. The above arguments seem to 
be specific enough to hope that a more complicated theory could 
avoid the presence of such matter.

\begin{figure}[!tb]
\centerline{\epsfxsize=13cm\epsfbox{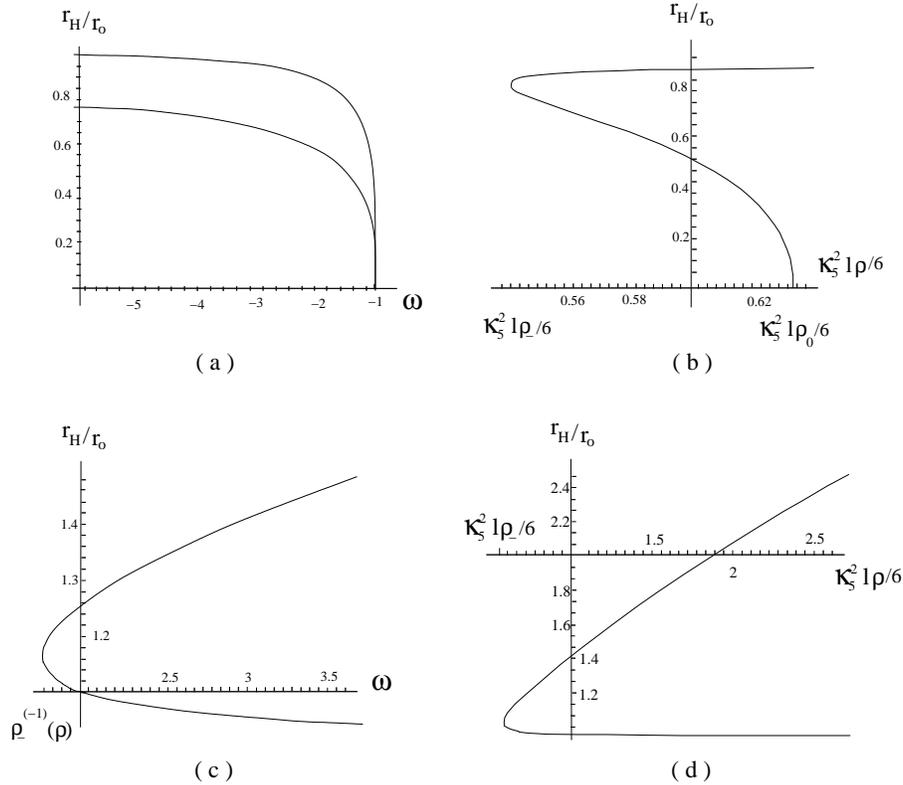}}
\caption[]{Displacement of the horizons when varying the energy density or the equation
of state on the brane. Both the inner and outer horizons are shown in
the plots.  (a) $\omega$ is varying in the region $\omega \leq -1$.
When $\omega$ reaches the critical point of a vacuum energy equation of state, the
two horizons hit the black hole singularity;
(b) the energy density is varying with $\omega<-1$ fixed. The two horizons
are between the black hole singularity and the brane. When $\rho$ goes
to $\rho_-$, the two horizons degenerate and when $\rho$ goes to $\rho_0$, the inner
horizon reaches the singularity;
(c) $\rho$ fixed, $\omega$ varying in the region $\omega>0$;
(d) $\omega>0$ fixed, $\rho$ varying. The two horizons belong to the cut region of the
BH space-time. When $\rho$ goes to $\rho_0$, the two horizons degenerate.
}
\label{fig:Horizons}
\end{figure}

So far we have only discussed solutions which include the black hole singularity
and cutting away, by the $\mathbb{Z}_2$ orbifold symmetry, the region close to infinity.
The motivation was to obtain an extra dimension of finite volume. The infinity
of the black hole space-time asymptotes to AdS space, just like
in a RS model with a single negative brane. This  would give a divergent
effective 4D Planck scale and therefore this possibility is
excluded. However,
as suggested in our previous analysis, the possibility to find a brane sitting
between the singularity and the two horizons brings another way to prevent
the 4D Planck scale to diverge: the singularity region is now cut away requiring a negative
energy density on the brane but the space naturally ends at the inner horizon so that
the 4D physics on the brane remains insensitive to the region beyond this inner
horizon. Even though the sign in the first jump equation
(\ref{eq:hjump}) flips,
when squared the jump equations result in the same relation
between the mass and the charge of the BH and the matter on the
brane\footnote{We thank Jim Cline and Hassan Firouzjahi for pointing out
a sign error in the first version of this paper, which lead to an
incorrect relation between the BH mass and charge, and the brane matter.}
\begin{equation}
	\label{eq:RN2muQ}
\mu = 3\left( l^{-2} +\sfrac{1}{36} \kappa_5^4 \omega \rho^2 \right) r_0^4,
\ \ \
Q^2 = 2\left( l^{-2} +\sfrac{1}{72} \kappa_5^4 (1+3\omega)\rho^2 \right) r_0^6.
\end{equation}
Repeating our previous analysis we conclude now that the horizons belong to the
part of the space that is kept {\it iff} $\omega \geq 0$.
The constraint for these horizons to exist then becomes
\begin{equation}
\rho<
-\sfrac{6 }{l\kappa_5^2}\sqrt{\sfrac{1}{8\omega^3}(1+6\omega-3\omega^2 +
\sqrt{(1+\omega)^3(1+9\omega)})}.
\end{equation}
Note that, even if now $\omega>0$, the second weak energy condition
$\rho+p>0$ is still violated.
Thus the situation is actually quite similar to the previous construction and
requires some exotic vacuum on the brane. Furthermore there is no way to
consider these solutions as perturbations around the RS setup since the allowed
region for the parameter $\omega$ is disconnected from the tension type 
equation
of state. In the remaining part of the paper, we concentrate on the first type
of solutions namely cutting the spacetime near infinity and keeping the
region that includes the singularity.

\subsection{Cutting the space at the horizon and the effective
cosmological constant}

We would like to show in this subsection that if we cut the space at the horizon,
the 4D effective vacuum energy, computed as the integral of the action over the
extra dimension, vanishes. This result may be expected because we were able
to construct a solution to the 
Einstein equations which is 4D Poincar\'e invariant
on the brane. However, unlike in other self-tuning solutions~\cite{ADKS,KSS}
where the space-time is cut at a naked singularity,
the cut at a horizon is natural
since from an observer's point of view it will take an infinite time
to reach the horizon and boundary conditions at a horizon are well-defined and
do not rely on the introduction of any {\it ad hoc} extra brane where some
fine-tunings are reintroduced (see \cite{FLLN}).

So we would like to show that the following quantity is indeed vanishing:
\begin{equation}
{\cal I} =
\int_{r_H}^{r_0} dr \sqrt{|g_5|} \left(
\sfrac{1}{2\kappa^5} R - \sfrac{1}{4} F^2 - \Lambda_{bk} +
{\cal L}_{\rm mat.} \delta (\sqrt{g_{rr}(r-r_0)}) \right)
\end{equation}
The first step is to evaluate the singular part coming from the brane namely to find
the expression of ${\cal L}_{\rm mat.}$ in terms of the 4D energy density and pressure
that model the brane. As we have emphasized before, for the case where a horizon is
present one needs an unconventional source with $\omega <-1$, which can not be 
obtained from stable dynamical fields, but should rather be thought of as 
a non-dynamical object. However, in order to calculate the relation between the
energy densities and the original Lagrangian we assume that the results for the
region $-1 < \omega < 1$ also apply for the case $\omega <-1$. In order
to calculate the relation we are interested in 
we assume that the matter corresponds to a
time-dependent scalar field
\begin{equation}
\label{scalar}
{\cal L}_{\rm mat.} = -\sfrac{1}{2} \partial_\mu \phi \partial^\mu \phi - V(\phi),
\end{equation}
from which we obtain that
\begin{equation}
\label{p}
{\cal L}_{\rm mat.} = p =
-\frac{h'(r_0)}{\kappa_5^2 \sqrt{h(r_0)}} - \frac{4\sqrt{h(r_0)}}{\kappa_5^2 r_0},
\end{equation}
where we have used the jump equations (\ref{eq:hjump}) to derive the last equality. 
We emphasize again that for the case of matter with $\omega <-1$ (as we have in the case
with a horizon) the assumption that the matter is dynamical like in (\ref{scalar}) would
imply an instability in the system, therefore one has to assume that rather
than being a dynamical matter it is more like a topological object, similar
to an orientifold. However, we assume that (\ref{p}) holds even in this
case.

In computing the bulk part of the integral, we must include the
singular part of the curvature at the $\mathbb{Z}_2$ orbifold
fixed point. To evaluate this singular contribution,
we can use
the form (\ref{eq:ABCmetric}) of the bulk metric
with $A(r)=\sqrt{h(r)}, B(r)=r$ and $C(r)=1/\sqrt{h(r)}$;
its curvature is given by~\cite{CG} (for $k=0$),
\begin{equation}
R = -\frac{2}{C^2}
\left( \frac{A''}{A} + 3 \frac{B''}{B}
+ 3 \frac{{B'}^2}{B^2} + 3 \frac{A'B'}{AB}
-\frac{A'C'}{AC} - 3 \frac{B'C'}{BC} \right)
=-h''(r)-6\frac{h'(r)}{r}-6 \frac{h(r)}{r^2}
.
\end{equation}
But because of the $\mathbb{Z}_2$ symmetry at $r=r_0$, the first derivatives
are discontinuous which translates into delta function type singularities
in the second derivatives. So the singular part of the curvature reads,\footnote{
The sign is fixed by keeping the region of space-time between the singularity
and the brane.}
\begin{equation}
R_{\rm sing.} =
\frac{4}{C^2} \left( \frac{A'}{A} + 3 \frac{B'}{B'} \right)
\delta (r-r_0)
=
\left( 2 h'(r_0) + 12 \frac{h(r_0)}{r_0} \right) \delta (r-r_0)
.
\end{equation}
And finally, using the expression (\ref{eq:AdSSch}) of the
bulk cosmological constant and the expression (\ref{eq:gauge})
of the gauge field strength,
the effective cosmological constant becomes
\begin{equation}
{\cal I} =
-\sfrac{1}{\kappa_5^2} \left[  l^{-2}r^4 + \frac{Q^2}{r^2} \right]_{r_H}^{r_0}
+\sfrac{1}{4\kappa_5^2} \left( 2 r^3 h' + 12 r^2 h \right)_{|r_0}
+\sfrac{1}{2\kappa_5^2} \left( -r^3 h' - 4 r^2 h \right)_{|r_0}
=
\kappa_5^{-2} r_H^2 h(r_H)
\end{equation}
which vanishes precisely because $r_H$ is the position of the horizon.
We can explicitly see that, as expected,  this cancellation does not require
adding  anything at the horizon.

\subsection{Maximally symmetric solutions on the brane}
\label{sec:MaxSym}

So far we have only considered flat brane solutions, but as we have discussed
in the Introduction, it is important to look for other 4D maximally symmetric
metrics on the brane, namely de Sitter or anti-de Sitter 4D space-time.
In the 4D adjustment mechanism proposed by Hawking~\cite{Hawking}, a four form
field provides a contribution to the cosmological constant whose
magnitude is not determined by the field equations but appears as
a constant of integration.\footnote{For a modern version of this idea in
a supersymmetric/superstring context, see~\cite{4Form}.}
It was argued that its probability distribution
may be exponentially peaked such that the flat solution will be preferred among
the other maximally symmetric solutions. Unfortunately, Duff has shown~\cite{Duff} that the vanishing of the effective cosmological constant is the most
unlikely possibility and the anthropic principle has to be invoked~\cite{HT} to
disentangle the different solutions. On the contrary, in the self-tuning
approach to the cosmological constant on a brane, the 4D Minkowski metric provides
the only maximally symmetric solution to Einstein's equations with an adequate
choice of the conformal coupling to the brane~\cite{ADKS}. 
We will argue now that self-selection
of the flat brane solution
remains true when the bulk
scalar field is replaced by a vector field, without tuning parameters in the 
action.

De Sitter and anti-de Sitter solutions can be parametrized
as FRW space-times,
\begin{equation}
ds^2 = -d\tau^2 + R^2(\tau) d\Sigma_k^2.
\end{equation}
The explicit form of the scale factor can be deduced from the fact
that (A)dS$_4$ space-time are solutions to the usual 4D Friedmann equation with
a (negative) positive vacuum energy.  This requires,
\begin{equation}
\frac{{\dot R}^2}{R^2} = \sfrac{1}{3} \kappa_4^2 \Lambda_4 -\frac{k}{R^2}
\end{equation}
Solving this ordinary differential equation leads to the
following parameterizations~\cite{HE,Kraus}:
\begin{eqnarray}
	\label{eq:dS}
&&
\mbox{de Sitter: }
\
\left\{
\begin{array}{lll}
k=0 &\mbox{ and } & R(\tau)=R_0 e^{H\tau};\\
k=-1 &\mbox{ and } & R(\tau)= \sinh (H\tau)/H;\\
k=1 &\mbox{ and } & R(\tau)= \cosh (H\tau)/H.\\
\end{array}
\right.
\\
	\label{eq:AdS}
&&
\mbox{anti-de Sitter: }
\begin{array}{lll}
k=-1 & \mbox{ and }  & R(\tau)=\cos (H\tau)/H.
\end{array}
\end{eqnarray}

If such solutions exist, they must correspond to moving branes in the
static bulk metric and thus must satisfy the jump equations
(\ref{eq:generaljump1})--(\ref{eq:generaljump2}). For a non-static brane,
the second jump equation is equivalent to energy conservation, so
$R(\tau)$  must satisfy,
\begin{equation}
	\label{eq:nonflat}
\frac{{\dot R}^2}{R^2} = \sfrac{1}{36}\kappa_5^4 \rho^2
-\left(\frac{k}{R^2}+\frac{1}{l^2}-\frac{\mu}{R^4}+\frac{Q^2}{R^6}\right)
\ \ \mbox{and } \ \
\dot \rho + 3 (1+\omega)\rho \frac{{\dot R}}{R} = 0 .
\end{equation}
When the parameter $\omega$ appearing in  the equation of state is constant,
the conservation equation becomes,
\begin{equation}
\rho = \rho_0 \left( \frac{R(\tau)}{R_0} \right)^{-3(1+\omega)}.
\end{equation}
Plugging back into the first differential equation in (\ref{eq:nonflat}), 
we determine
in which cases the (A)dS space-time are solutions to the jump equations
\begin{itemize}
\item The 4D de Sitter metrics will be solution only when the equation of state
on the brane corresponds to a vacuum energy, $\omega=-1$, and
when this vacuum energy
is bigger than the contribution from the bulk, in which case
the Hubble parameter is given by $H^2=\sfrac{1}{36}\kappa_5^4\rho^2-l^{-2}$.
All these solutions are moving in a pure Anti-de Sitter bulk, namely
$\mu=0$ and $Q^2=0$.
\item The 4D Anti-de Sitter metrics will be solutions for a vacuum energy
type brane ($\omega=-1$) with small enough energy density, in which case the 
Hubble parameter
is given by $H^2=l^{-2}-\sfrac{1}{36}\kappa_5^4\rho^2$ while the
charge and the mass still vanish. These solutions are
the AdS$_4$ branes in AdS$_5$ bulk studied in~\cite{KR}.
There are other 4D AdS solutions
where the Hubble parameter 
is given by the 5D vacuum energy, $H^2=l^{-2}$, and the energy
density on the brane is balanced respectively by the (negative)
mass of the 5D black hole if $\omega=-1/3$
($\mu=-\sfrac{1}{36}\kappa_5^4 l^4\rho_0^2$ and $Q=0$)
or by its charge if $\omega=0$
($Q^2=\sfrac{1}{36}\kappa_5^4l^6\rho^2_0$ and $\mu=0$).
\end{itemize}
This analysis shows that the 4D Minkowski metric is the only maximally symmetric
solution on the brane up to, for particular equations of state,
a discrete choice of the constants of integration.  
Hence, self-tuning in these models is somewhat natural: for a generic brane
equation of state, the only continuous class of solutions is the set of
flat brane solutions. 
This result offers a selection of a vanishing cosmological constant from
symmetry requirements only. The vanishing of the cosmological constant
is ensured by an adjustment of the charge and the mass of the black hole
through gravitational waves emitted from the brane when a phase transition
occurs for instance. A precise description of such a phase transition
and the response of the bulk would however certainly deserve further scrutiny.

\section{Lorentz violations in black hole backgrounds}
\setcounter{equation}{0}
\setcounter{footnote}{0}

We have seen in the previous section that asymmetrically warped
spacetimes due to black holes in the bulk can exist, and perhaps even provide
an adjustment mechanism for the effective cosmological constant. In this 
section, we investigate some of the physical consequences of such 
spacetimes. In this analysis we will not assume that the cosmological constant
problem is resolved by these backgrounds, but we are rather interested in 
general consequences of having an asymmetrically warped metric. In particular,
we will not need to have exotic energy densities on the branes (for which the
price to pay is the reintroduction of fine-tuning in the theories).
First we calculate the speed of gravitational waves based on
the analysis of lightlike geodesics, then show how the graviton zero mode
found by Randall and Sundrum would be modified in such spacetimes. 
Finally, we consider the cosmology of these models in order to gain
some insight into the effective gravity observed on the brane.
 
\subsection{Geodesics in the black hole background and the speed of
gravitational waves}

One of the most important consequences of an asymmetrically warped spacetime
is that the speed of gravitational waves could differ from the speed of
light. The reason for this is that in asymmetrically warped 
metrics the local speed of light is a function of the coordinate 
along the extra dimension $r$:
\begin{equation}
c(r)=\left( \frac{h(r)\,l^2}{r^2}\right)^{1/2}.
\end{equation}
Thus one might think of this spacetime as a medium with a continuously
changing index of refraction $n(r)$, in which the propagation of light is
 the analog of the propagation of gravitational waves 
in the asymmetrically warped spacetimes. Propagation of light is 
governed by Fermat's principle, and if $c(r)$ is increasing away from
the brane it may be advantageous for gravitational waves to bend into
the bulk, and arrive earlier than waves propagating along the
brane would. However since electromagnetism is forced to propagate 
along the brane, it will always keep the local velocity at the brane
$c(r_0)$. Thus we expect as long as the velocity $c(r)$ increases
away from the brane gravitational waves can travel
faster than the local speed of light at the brane. This will lead to an 
apparent violation of
Lorentz invariance due to the bulk dynamics in the low energy effective 
theory, and an apparent violation
of causality from the brane observer's point of view,
in the sense that gravitational waves emitted from a source will arrive
earlier than light signals from the same source. We should emphasize that
this is not a ``real'' violation of causality. Since there are no closed 
timelike curves in the theory, propagation of massless particles always 
proceeds forward in time. Apparent violation of causality simply means that
the regions of spacetime which are in causal contact are  different from what 
one would naively expect from an ordinary Lorentz invariant 4D theory.
This is an 
experimentally verifiable prediction for these models, and can be 
tested by gravitational wave experiments like LIGO, VIRGO  or LISA.
Some of these points were made independently in 
Refs.~\cite{Kraus,Kalbermann,CF,Youm,Ishihara,CKR}
and it has been argued that
this apparent Lorentz symmetry violation may provide an alternative to inflation
to deal with the horizon problem.  This is similar to 
another alternative to solve the horizon
problem proposed in~\cite{Albrecht,superluminal,Kiritsis}, where
the speed of light is changing in time, thus regions of space-time
may have been in causal contact even if they appear to be outside each others
horizons. Violations of Lorentz
invariance similar to the ones considered in this paper
may also appear in non-commutative gauge theories~\cite{LLT,Bak}.
In generic non-commutative theories faster than light propagation
will not be suppressed in the perturbative sector, and can not give
a realistic model. In supersymmetric theories however the Lorentz violation
in the perturbative sector vanishes, and will appear only as faster-than-light
propagation in the solitonic sector, which is very weakly coupled to
the perturbative states. This way interesting theories~\cite{Aki}
analogous to the brane constructions presented here can be obtain
from non-commutative gauge theories. However, it remains to be seen
whether or not supersymmetry breaking will reintroduce the Lorentz breaking
effects to the perturbative sector.

If $c(r)$ is a decreasing function as one moves away from the
brane, for example in the case that there are horizons shielding the
singularity from the brane,
we expect that those gravitational waves that will be observed
are those that remain on the brane, 
and thus no discrepancy between the speed of light and the speed of gravity
should exist.  It would be interesting to study the graviton wave equation
in these cases, which we postpone untill the next subsection.
Below we show, that the analysis of lightlike geodesics
is in complete agreement with the physical intuition sketched above.

Here we analyze the geodesics of the theories with asymmetrically warped 
metrics. Some of the features of geodesics in asymmetrically warped 
spacetimes have also been analyzed in~\cite{Kalbermann,CF,Ishihara}.
The equations for the geodesics are given by
\begin{equation}
	\label{eq:geo}
\frac{d^2 x^M}{d\tau^2} + \Gamma^M_{PQ} \frac{d x^P}{d\tau}\frac{d x^Q}{d\tau} =0,
\end{equation}
where $\tau$ is an affine parameter along the trajectory (the proper time
for the spacelike geodesic). As usual, these equations can be deduced
from the Lagrangian provided by the proper distance. In the case of the black 
hole metric:
\begin{equation}
ds^2 = -h(r)\,dt^2 + l^{-2}r^2\,d \vec{x}^2 + h(r)^{-1}dr^2,
\end{equation}
the Lagrangian is explicitly given by:
\begin{equation}
{\cal L} = \frac{ds^2}{d\tau^2} =
-h(r)\,\dot{t}^2 +  l^{-2}r^2\,\dot{\vec{x}}^2  +  h^{-1}(r)\,\dot{r}^2,
\end{equation}
where the dot represents differentiation with respect to the affine 
parameter $\tau$.
Furthermore the Euler--Lagrange equations are supplemented by
the consistency condition defining the affine parameter:
\begin{equation}
{\cal L} = \epsilon, \ \ \
\epsilon=-1/0/+1 \mbox{ for timelike/lightlike/spacelike geodesics, respectively.}
\end{equation}

From the $t$- and $\vec{x}$-independence of the black hole metric,
we find four Killing vectors and thus four corresponding conserved quantities:
\begin{eqnarray}
&&
\frac{\partial{\cal L}}{\partial \dot{t}}
= -2h(r)\,\dot{t} = {\rm const.} \equiv -2E, \\
&&
\frac{\partial{\cal L}}{\partial \dot{x^i}}
= 2l^{-2} r^2 \,\dot{x^i} = {\rm const.} \equiv 2 p^i.
\end{eqnarray}
In terms of these conserved quantities, the consistency equation is
\begin{equation}
	\label{eq:consistency}
{\dot r}^2 + \frac{h(r)}{r^2}\, l^2 \vec{p}^2 - h(r) \epsilon  = E^2.
\end{equation}
As long as $ {\dot r} \not = 0$, this equation is equivalent to the $r$ 
component
of the geodesic equations (\ref{eq:geo}). However, for a straight geodesic 
parallel
to the brane, {\it i.e.}, ${\dot r}=0$, the first derivative of 
(\ref{eq:consistency}) also has 
to be satisfied.

From now, we will concentrate on lightlike geodesics only.
The motion in the $r$ direction transverse to the brane is then analogous
to Newtonian dynamics for a particle of two units of mass
with energy $E^2$ moving in the potential
\begin{equation}
V(r) = \frac{h(r)}{r^2}\, l^2 \vec{p}^2 .
\end{equation}
The behavior of V(r) around the brane in the direction of the singularity,
and the position of the brane with respect to the horizons if they exist,
will determine the shape of the geodesics in the bulk.

\begin{figure}[!tb]
\centerline{\epsfxsize=16cm\epsfbox{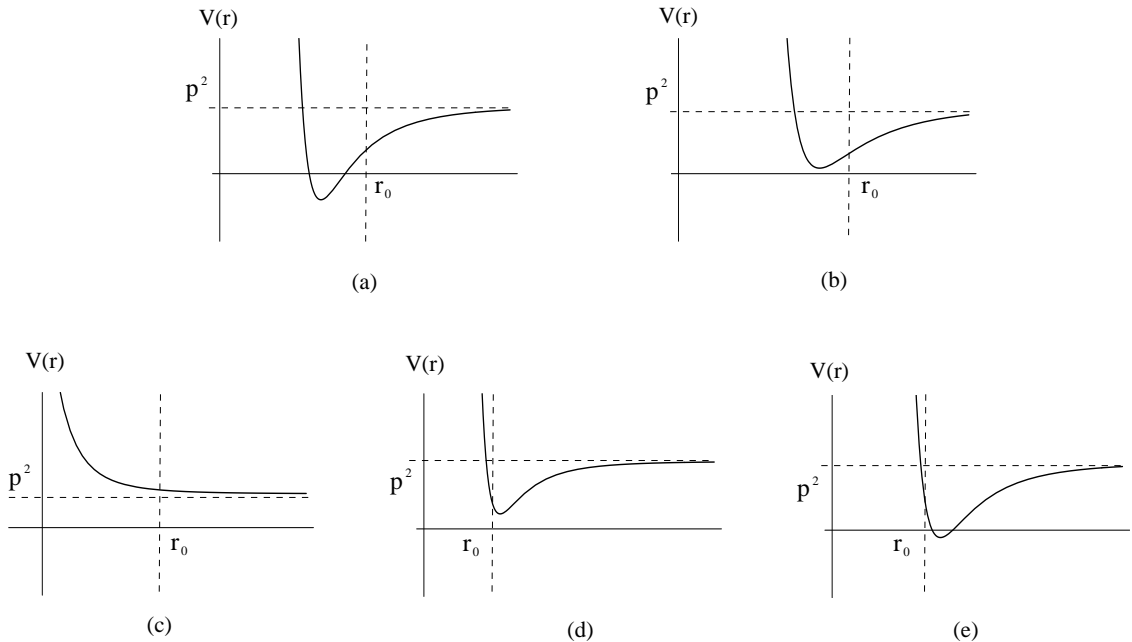}}
\caption[]{Shapes of the Newtonian potential. The functions $\rho_-, \rho_0$
and $\rho_\mu$ defined in (\ref{eq:rho_pm}), (\ref{eq:rho0}) and
(\ref{eq:rhoMu}) border the different regions in the plane $(\omega,\rho)$.
The asymptotic value of the potential at infinity corresponds to the
momentum $|\vec{p}|^2$ along the brane. The space is however cut at the
brane sitting at $r=r_0$. We identify five different shapes for the potential.
(a) $\omega<-1$ and $\rho_0>\rho>\rho_-$: the potential vanishes at the
two horizons. The geodesics will come back to the brane but, from a brane
observer's point of view, it already takes an infinite time to cross the 
horizon and
so the geodesics will never be seen as coming back.
(b) $\omega<-1$ and $\rho_->\rho$. The 4D local speed of graviton
starts decreasing when the geodesic leaves the brane.
(c) $-1<\omega<0$ and $\rho>\rho_\mu$.
(d) $-1<\omega<0$ and $\rho_\mu>\rho$ or $0<\omega$ and $\rho_->\rho$.
(e) $0<\omega$ and $\rho>\rho_-$.
In the three last configurations, the 4D speed of the graviton is
increasing when going into the bulk, and on its return the geodesic may
reach a
point on the brane where light emitted with the graviton
has not yet arrived.
}
\label{fig:pot}
\end{figure}

From the jump equations (\ref{eq:RNmuQ}), the expression for the potential
becomes
\begin{equation}
V(r) = \vec{p}^2 \left( 1-{\hat \mu} r_0^4/r^4 +{\hat Q}^2 r_0^6/r^6 \right),
\end{equation}
with
\begin{equation}
{\hat \mu}= l^2 \mu /r_0^4=
-3\left( 1 +\sfrac{1}{36} \kappa_5^4 \omega l^2 \rho^2 \right),
\ \
{\hat Q}^2 = l^2 Q^2 /r_0^6=
2\left( 1 +\sfrac{1}{72} \kappa_5^4 (1+3\omega)l^2 \rho^2 \right).
\end{equation}
The shape of the potential $V(r)$ will depend on the sign of
${\hat \mu}$ and whether there exists a minimum between the singularity and the brane.
The condition for ${\hat \mu}$ to be positive is
\begin{equation}
	\label{eq:rhoMu}
\left( \omega>0 \right)
\ \ \ \mbox{ or } \ \ \
\left( \omega<0 \mbox{ and } \rho<\rho_\mu = \sqrt{\sfrac{-36}{\omega l^2\kappa_5^4}}
\right).
\end{equation}
And when ${\hat \mu}>0$, the minimum of the potential will be at
\begin{equation}
r=r_0 \sqrt{\frac{3 {\hat Q}^2}{2{\hat \mu}}}.
\end{equation}
To locate the position of this minimum with respect to the brane, we can
study the ratio ${\hat Q}^2/{\hat \mu}$, or equivalently, compute the {\it force},
$-V'(r)$, felt by the analogous Newtonian particle at the position of the brane:
\begin{equation}
-V'(r_0)=\frac{\vec{p}^2}{r_0} \left(-4 {\hat \mu} + 6 {\hat Q}^2 \right)
=\frac{\vec{p}^2}{6 r_0} (1+\omega) l^2 \kappa_5^4 \rho^2,
\end{equation}
from where we conclude that the minimum will be between the brane and the singularity
{\it iff} $\omega<-1$. In summary, five different shapes for the potential arise and they are drawn in
Fig.~\ref{fig:pot}. This also shows that based on the intuitive
picture presented at the beginning of this section one would expect the
speed of gravitational waves can be larger than the speed of light 
when $\omega \geq -1$, since this is the case when the speed of light
away from the brane increases. 

\begin{figure}[!tb]
\centerline{\epsfxsize=16cm\epsfbox{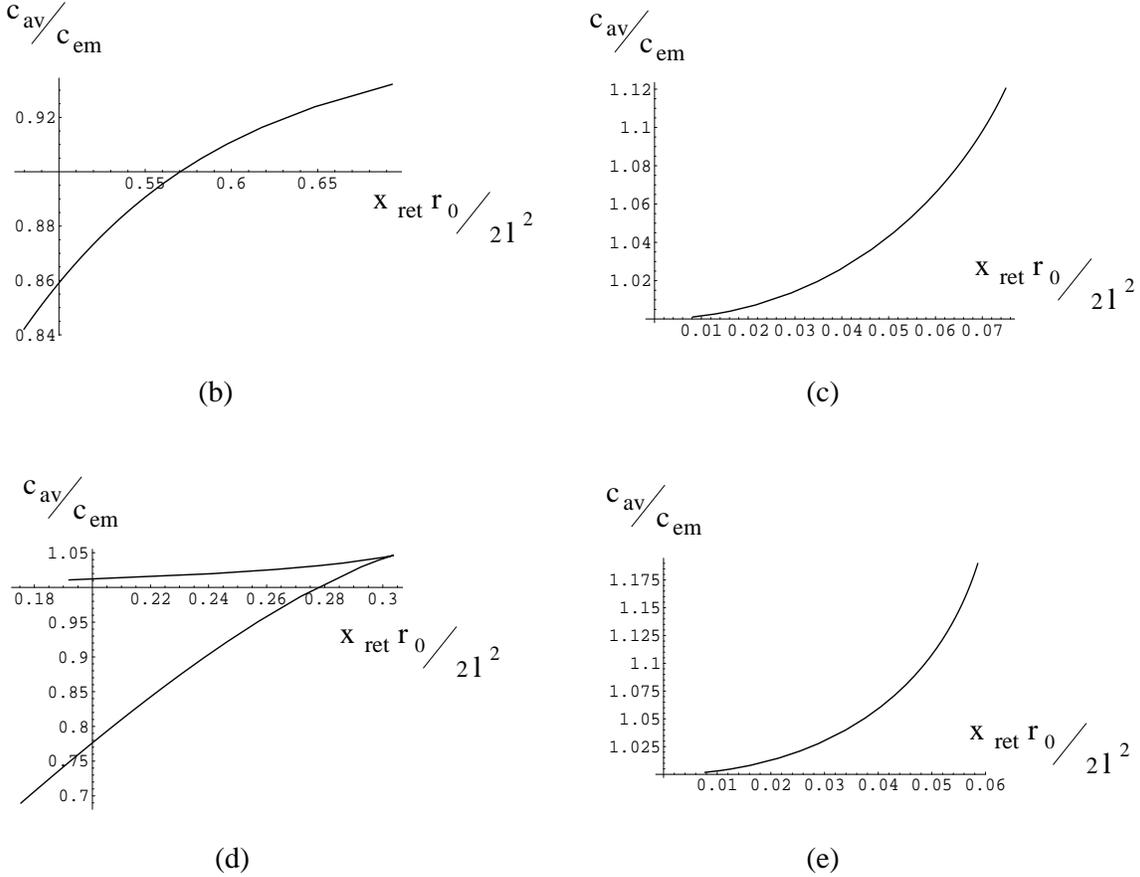}}
\caption[]{The Average speed of gravitons propagating along a geodesic off the
brane as function of the distance on the brane. We clearly see that in the region
of the plane $(\omega,\rho)$ where the Newtonian potential behaves like
the shapes (c),(d) and (e) of Fig.~\ref{fig:pot}, the graviton can propagate
faster than the light and its speed increases with the distance to the source on the
brane. We also note in the case (d) that there are various ways for gravity
to propagate between the same points on the brane, {\it i.e.} different
geodesics in the bulk with different values of $E/|\vec{p}|$ can return
to the same point on the brane. When $E/|\vec{p}|$ is too large, then the 
average speed
becomes lower than the speed of light.
In the case (b), the graviton will always propagate along the
brane with a speed faster than that in the bulk.
}
\label{fig:averagespeed}
\end{figure}

Finally, we calculate the average speed of gravitational waves as observed
from the brane observer's point of view. This can be calculated for the
case when the speed of light away from the brane increases by first 
calculating the turning point of the Newtonian particle in the 
potential $V(r)$, which is the largest solution\footnote{Note that
we have cut the space at the brane and kept the region between the singularity
and the brane. Since the potential diverges at the origin, there always exists
a point where $\dot{r}=0$ at which the potential is equal to the energy.
This solution corresponds to the turning point where the
geodesic starts moving back to the brane.} between
the singularity ($r=0$) and the brane ($r=r_0$)
of the following equation:
\begin{equation}
\left(\frac{E}{|\vec{p}|}\right)^2  = \frac{h(r_T)}{r_T^2} l^2.
\end{equation}
Once the turning point is obtained (numerically), one can eliminate the
proper time by dividing the expressions for $\dot{x}$ by the expression
for $\dot{r}$, and integrate the resulting equation to obtain $x(r)$.
With this the distance on the brane $x_{ret}$ after which the geodesic returns
to the brane can be expressed as
\begin{equation}
x_{ret} \left( E/|\vec{p}| \right)
=
\int_{r_T}^{r_0}
\frac{2\, l^2 \, dr}{r^2 \sqrt{\left( E/|\vec{p}|\right)^2 - h(r) l^2 / r^2}},
\end{equation}
and similarly the time it takes to return can be expressed as
\begin{equation}
	\label{average}
t_{ret} \left( E/|\vec{p}| \right)
=
\int_{r_T}^{r_0}
\frac{2\, E/|\vec{p}| \, dr}{h(r)\sqrt{\left( E/|\vec{p}|\right)^2
-  h(r) l^2 / r^2}}.
\end{equation}
To obtain the relative speed compared to the speed of light,
the average speed $c_{av}=x_{ret}/t_{ret}$ has to be compared with 
the local speed of light at the brane
$c_{em}=( h(r_0)l^2/r_0^2)^{1/2}$.
One can clearly see from (\ref{average}) that
the average speed evaluated from the geodesics will depend on the
value of $E/|\vec{p}|$, which is equivalent to choosing the oscillation length
of the gravitational wave around the brane (which is also equivalent
to how far into the bulk the gravitational wave is penetrating). The
dependence of the average speed on the oscillation length is given in
Fig.~\ref{fig:averagespeed} for the cases where the speed of light
away from the brane is increasing (that is cases (c), (d) and (e)) of
Fig.~\ref{fig:pot}. The full gravitational propagation would be
given by a superposition of these simple oscillating modes.

Depending on the exact structure of the low-energy effective theory,
the Lorentz violation due to bulk effects may be transmitted to particles on 
the 
branes by gravitational loops. At the scale where gravity becomes strongly
interacting, {\it i.e.} at the fundamental Planck scale, $M_*$, such loops 
will not be suppressed.
In order for particle physics to remain Lorentz invariant, the Lorentz violations
should decrease as the energy scale is increasing, otherwise one would expect
unsuppressed Lorentz violating operators due to gravitational loops. One can see
that in most cases considered here such constraints can be generically satisfied.
The reason can be understood from a holographic argument, similar to
ones explained in~\cite{holoRGE}. As the energy scale is increasing, gravity on the
brane is probing less and less of the bulk region around the brane,
indeed gravitational waves simply will have less time to travel
further into the bulk. Thus in order to avoid large Lorentz
violating operators being generated through gravitational loop effects
one has to arrange that
the region around the brane at a distance of the order $M_*^{-1}$ should be 
very close to ordinary AdS space. Of course this can always be achieved by moving the
black hole far away from the brane. In the language of the geodesic analysis 
of this section this constraint would imply that at small distances the 
speed of gravitational waves should approach the speed of light on the 
brane. One can see that the cases (c) and (e) 
in Fig.~\ref{fig:averagespeed} automatically satisfy this 
requirement, and thus by adjusting $r_0$ one can satisfy experimental constraints
on Lorentz violations in particle physics.\footnote{We thank John Terning for 
discussions on this issue.}

%

\subsection{A perturbative zero mode}

We have seen from the analysis of the geodesics that one expects gravity
to propagate with a speed different from ordinary electromagnetism for
asymmetrically warped spacetimes. This can cause apparent
violations of causality and Lorentz invariance in the gravitational
sector, without affecting particle physics (except through gravitational 
loops).

Next we show an analysis different from the geodesic approach to 
demonstrate the same effect. We will examine how the graviton zero mode
of the Randall-Sundrum model is modified in one of the black hole 
spacetimes considered in this paper. To simplify the equations, we will
consider the black hole metric as a linearized perturbation around
the RS spacetime. We have chosen to analyze the perhaps simplest 
form of matter on the brane, the case which simply corresponds to
a brane tension ($\omega = -1$). As we have seen before, in this case 
there is no horizon, but there is a naked singularity away from the 
brane. In order to be able to analyze this theory as a perturbation
around the RS metric, we have to cut the space-time with a second
(fine-tuned) brane before
the deviation from the RS spacetime becomes large (and of course before
the appearance of the naked singularity). This way we obtain a metric
that is close to the RS metric everywhere, and we will think of the
mass and charge of the black hole as a perturbative expansion 
around the RS solution. In order to make this expansion more
transparent, we first transform the black hole metric
by an appropriate rescaling of the coordinates $t$ and $\vec{x}$ and
a coordinate transformation $r=r_0 \exp(-k y)$, where $k=1/l,$ to the
form
\begin{equation}
	\label{eq:RStype}
ds^2=-e^{-2 k |y|} \hat{h}(y)\, dt^2
+ e^{-2 k |y|} d\vec{x}^2 +
\hat{h}(y)^{-1} \, dy^2 ,
\end{equation}
with $\hat{h}(y)=1-\mu l^2 r_0^{-4} e^{4 k y} + Q^2 l^2 r_0^{-6} e^{6 k y}$.
The location of the brane is now at $y=0$. As stated above, we also assume that
$\mu l^2 r_0^{-4} e^{4 k y}$ and $Q^2 l^2 r_0^{-6} e^{6 k y}$
remain small everywhere in the bulk, and thus $\hat{h}(y)^{-1}$
can be expanded in $\mu$ and $Q^2$. Now we would like to solve
the linearized Einstein equation in this background, and in
particular
find the modified propagation speed of the graviton zero mode.
One can show that the transverse traceless modes of the graviton $h_{\mu\nu}$
satisfy the following linearized equation:
\begin{eqnarray}
&&
-\sfrac{1}{2} D_M D^M h_{\mu\nu}
+\sfrac{1}{2} R_{M\mu}\, h_\nu{}^M
+\sfrac{1}{2} R_{M\nu}\, h_\mu{}^M
-R_{\mu\rho\nu\sigma}\, h^{\rho\sigma}
+\sfrac{1}{2} R_{\rho\sigma}\, h^{\rho\sigma}\, g_{\mu\nu}
-\sfrac{1}{2} R\, h_{\mu\nu}
\nonumber\\
	\label{linearized}
&&
\ \ \ \
=
\sfrac{1}{2}\kappa_5^2 F_{\rho M} F_\sigma{}^M h^{\rho \sigma} g_{\mu\nu}
-\sfrac{1}{4}\kappa_5^2 F_{MN} F^{MN} h_{\mu\nu}
-\kappa_5^2 \Lambda_{bk}\, h_{\mu\nu},
\end{eqnarray}
where the covariant derivatives are with respect to the background metric
$g_{MN}$, as are the Riemann tensor and the Ricci tensor and scalar.
We have
also checked that these transverse traceless
modes decouple from perturbations of the gauge field
and thus only the background gauge field appears in the right
hand side of the equation. In fact, one can
show that after explicitly including the expressions for the covariant
derivatives and the background quantities in (\ref{linearized})
the equation simply becomes identical
to the equation for a minimally coupled massless scalar in the bulk, just
like in the case of the RS-type backgrounds~\cite{Universal}.
Hence,
we simply study the propagation of a scalar field $\Phi$ in the bulk.
We know that for vanishing $\mu$ and $Q^2$ the zero mode solution 
in the RS model is just $\Phi=\Phi_0 e^{i(\omega t-\vec{q}\cdot \vec{x})}$,
where $\omega^2=q^2$, and $\Phi_0$ is a normalization
constant. Thus we look for a perturbative solution of the
bulk equation of the form,
\begin{equation}
\Phi=
\Phi_0 (1+\delta \Phi (y)) e^{i\left((\omega +\delta \omega )t -\vec{q}\cdot \vec{x}\right)},
\end{equation}
where $\delta \Phi (y)$ and $\delta \omega$ are assumed to be proportional
to $\mu$ and $Q^2$. A scalar zero mode of a similar
form has been found for another asymmetrically warped metric in
Ref.~\cite{CKR}.
Expanding the linearized equation around the background (\ref{eq:RStype})
one obtains,
\begin{equation}
\delta \Phi'' - 4 k \delta \Phi'
+\mu l^2 r_0^{-4} q^2 e^{6 k y}
- Q^2 l^2 r_0^{-6} q^2 e^{8 k y}
+ 2 \delta \omega\, q e^{2 k y}=0.
\end{equation}
As explained above, we are assuming for the sake of perturbativity that
the space-time is made finite by the introduction of a 
second brane, and does not include the singularity.
Therefore we need to impose a boundary condition on
$\delta\Phi$ that its derivative at the two branes vanishes, {\em i.e.}
\begin{equation}
\delta \Phi'|_{y=0,y_R}=0,
\end{equation}
where $y_R$ is the position of the regulator brane.
Enforcing these boundary conditions one obtains the dispersion for the zero
mode, which is given by
\begin{equation}
\delta \omega =
\sfrac{1}{4} \, q \left(
-2 \mu l^2 r_0^{-4} + Q^2 l^2 r_0^{-6} (1+e^{2 k y_R})
\right)  e^{2 k y_R} .
\end{equation}
This dispersion relation remains linear such that the speed of propagation
of gravitational waves is constant and given by
\begin{equation}
c_{grav} =
1+ \left(
-\frac{\mu l^2}{2r_0^4}  + \frac{Q^2 l^2}{4r_0^6} (1+e^{2 k y_R})
\right) e^{2 k y_R} ,
\end{equation}
which has to be compared with the propagation speed of electromagnetic waves
given by $c_{em}=\sqrt{\hat{h}(y=0)}=1-\sfrac{1}{2} \mu l^2 r_0^{-4}
+ \sfrac{1}{2} Q^2 l^2 r_0^{-6}$ in these coordinates. For the
$\omega=-1$ spacetimes $\mu$ and $Q$ are related by the formula
$\mu= \sfrac{3}{2} Q^2/r_0^2$, therefore the
difference of the two speeds can be written as
\begin{equation}
c_{grav}-c_{em}=
\left(\cosh (2ky_R)-1\right) e^{2ky_R} \frac{Q^2 l^2}{2 r_0^6}
\geq 0,
\end{equation}
Therefore the gravitational speed for the propagation of the zero mode
is always larger than the speed of
electromagnetic waves in the scenario considered here. The LIGO experiment
may be able to detect gravitational waves from type II supernovae up to a 
distance \cite{LIGO} of about 20 Mpc 
($\sim 6\cdot 10^7$ ly). For objects of such a
distance even a tiny difference in the speeds of gravitational and 
electromagnetic waves would cause a huge time difference, and thus the 
possible values of $\mu$ and $Q$ could be severly constrained \cite{CKR}.
In fact, the limitations of such measurements are likely not to lie in the
time resolution of the gravitational
and the light signal, but rather the opposite problem: if there is an
appreciable difference in the propagation speeds then due to the huge
distance to the expected sources the arrival time differences could turn
out to be way too big to be able to identify the fact that the source
for the gravitational wave and the light was the same. For a supernova 20
Mpc away from us, and very conservatively assuming that the arrival time
difference should be less than 5 years, in order to be able to actually
detect the different arrival times one needs to have the difference in
the speeds to be less than $\frac{\delta c}{c} \leq 10^{-7}$. Otherwise
the gravitational wave experiments will simply not be able to 
identify the source for the observed gravitational waves. Type I 
supernovae could likely be detected by LIGO only if they happen within our
galaxy. These are very rare, however assuming the best possible scenario
one could see a supernova a few hundred thousand light years from us. In this
case (again assuming a very conservative time difference of 5 years) 
the maximum value of $\frac{\delta c}{c}$ that could be tested is 
of the order of $10^{-3}$.

\subsection{Cosmological Expansion and Effective Gravity Theory}

So far we have discussed the interesting physical consequences of
asymmetrically warped spacetimes: the possible difference in the
speeds of gravitational and electromegnetic waves, 
and the possible adjustment of the cosmological constant
to give  a flat brane. Next we would like to understand what kind
of gravitational theory a 4D observer on the brane would see.
There are at least two distinct possibilities:\footnote{We thank Nemanja Kaloper
for discussions on this issue.}
the effective action for gravity could explicitely break the 4D Lorentz
invariance for instance by introducing different coefficients for time
and space derivatives in the kinetic terms of the graviton, or
the effective action may only violate 
the weak equivalence principle by the presence of
some extra fields which couple differently
to matter and gravitation forcing
the graviton to propagate differently 
as the other gauge bosons (see~\cite{Will} for a review
on the different tests of Einstein gravity). Note that the distance of the 
brane to the singularity or the horizon could be such a field.
Both approaches will manifest themselves experimentally by observing
different speeds of propagation
for the graviton and the photon.
Instead of trying to understand this (very important) question of how
to incorporate the Lorentz violating effects into the low
energy effective action,
we solve the simpler problem (which still gives us some insights into the
long distance behavior of gravity)
of finding the cosmological evolution of the brane  in the presence of
matter perturbations on top of the vacuum energy. We will express the
sources on the brane as the sum of the (non-dynamical) vacuum energy,
plus the matter perturbations:
\begin{equation}
\rho_{tot}=\rho_0+\rho, \ \ p_{tot}=p_0+p.
\end{equation}
The expansion of the brane can then be read off from
(\ref{jump1})-(\ref{jump2}). 
Using (\ref{RNjump}) the form of the expansion equation can be simplified to
\begin{equation}
	\label{FRW}
\left( \frac{\dot{R}}{R}\right)^2=\frac{\kappa_5^4 \rho_0}{18} \rho
+\mu \left( \frac{1}{R^4}-\frac{1}{r_0^4} \right)
-Q^2 \left( \frac{1}{R^6}-\frac{1}{r_0^6} \right)
+\frac{\kappa_5^4}{36} \rho^2,
\end{equation}
while the conservation of energy equation is given by the
slightly unconventional expression
\begin{equation}
\dot{\rho}+3 \left( \rho_0 (1+\omega_0)+\rho (1+\omega ) \right)
\frac{\dot{R}}{R} = 0.
\end{equation}
In these equations, $R$ is nothing but
the scale factor on the brane and a dot denotes
a derivative with respect to the proper time on the brane, {\it i.e.}, the usual
FRW time; $r_0$ is the position of the brane before matter was introduced.
On comparison with similar terms in usual 4D Friedmann equation,
from (\ref{FRW}) we can identify the 4D effective
Planck scale as 
\begin{equation}
\frac{1}{M_{Pl}^2}=\frac{\kappa_5^4 \rho_0}{6}.
\end{equation}
If one assumes that energy densities on the brane are of the order of the
TeV scale, then the required size of the five dimensional Planck scale
would be given by $M_*=10^8$ GeV, with $\kappa_5^2=1/M_*^3$.
One can also see that very close to the static solution (that is 
at $R=r_0$ there is no correction to the ordinary FRW equation, which
one could perhaps interpret as the zero mode of this model reproducing ordinary
4D gravity. However, as the brane moves further away from the static 
point, the corrections to the Friedmann equation will start becoming
sizeable. This has one positive consequence: a cosmological constant 
term $\mu/r_0^4-Q^2/r_0^6$, the usual dark radiation
term $\mu/R^4$ and due to the charge of the black hole
a contribution that is similar to the contribution of 
a 4D massless scalar, $Q^2/R^6$ is obtained,
which can be used to fit the Friedmann equation to the observed accelerating
Universe\cite{accelerating}.
This possibility has also been pointed out by Refs.~\cite{Visser,CoHo}.
However, these terms also pose a problem: if $\mu, Q^2$ and $r_0$ are
such that the cosmological constant gets adjusted to zero for $R=r_0$,
then after just a short expansion period the above mentioned 
new terms in the Friedmann equation will start dominating, if $\mu$ and $Q^2$
remain constant. In order to overcome this problem, one has to assume 
that the adjustment mechanism that sets the cosmological constant 
dynamically to zero also operates during the ordinary expansion of the 
Universe, thus making $\mu$ and $Q^2$ time dependent.  Let us
determine how small the characteristic time scale for the adjustment
mechanism would have to be in order for the solution to track the
vanishing cosmological expansion solution. In such a case the 
corrections due to the change in $\mu$ of the form $\dot{\mu}/R^4$
should cancel the corrections from the change in the position of the 
brane of the form $\mu\dot{R}/R^5$, (up to remaining terms
in the Friedmann equation of order $H^2=(\dot{R}/R)^2)$. 
This would lead to $\dot{\mu}/\mu \sim \dot{R}/R \sim H$;
thus, the characteristic time scale for the adjustment should be of the
order or shorter than the Hubble scale. If one wants an ordinary FRW Universe
after nucleosynthesis then one should require that this scale is shorter
than $H_{BBN}$. This is not a very strong restriction for the time scale 
of adjustment, and could still leave the possibility for early inflation
open. Of course if one is not trying to solve the cosmological constant problem,
then $\mu, Q^2$ and $r_0$ should be viewed as free parameters that are not
necessarily related to the fundamental Planck scale. In this case these parameters
can be simply used to fit the observed accelerating Universe without assuming any 
time dependence for them. In that case the acceleration of the expansion of the
Universe would be a manifestation of gravitational Lorentz violations in extra
dimensions rather than a consequence of a tiny bare vacuum energy in four dimensions.

\section{Conclusions}

The work of Randall and Sundrum has revealed that new physics can emerge
from a non\-trivial (``warped'')
geometry mixing the four dimensions of our brane-world
with an extra non-compact dimension.
While the main attention has been focused
on solutions preserving 4D Lorentz symmetry in the bulk,
other solutions exist with different warp factors for timelike and spatial
directions. In fact these {\it asymmetric} solutions are the more 
generic ones, 
and they open up 
new perspectives to low energy gravitational interactions
due to the Lorentz symmetry violation they can mediate.

One of the most striking
features is the possibility for gravitational waves to
propagate faster than electromagnetic waves stuck
on the brane~\cite{Kalbermann,CF,Ishihara,CKR}. This apparent
violation of causality from a brane observer point of view could be 
experimentally tested by gravitational wave detectors.
In the same vein, we have shown that the addition of a vector field in the bulk
alleviates the fine-tuning of the cosmological constant problem. The bulk
has a geometry of an AdS--Reissner--Nordstr\"om black hole, where the
relaxation could be achieved by the adjustment of the mass and charge 
of the black hole. A horizon could protect the black hole singularity,
however the existence of a horizon
requires the existence of some exotic energy density on 
the brane. The cosmology in asymmetrically warped spacetimes can explain the observed 
acceleration of the Universe which thus would appear
as a manifestation of gravitational Lorentz violations in extra dimensions.
If one wants to solve the cosmological constant problem
and explain the accelerating Universe simultaneously, then one has to assume that 
the relaxation of the mass and charge 
is of the order or faster than the Hubble scale.

\section*{Acknowledgments}
We thank Nima Arkani-Hamed, Francis Bernardeau,
Tanmoy Bhattacharya, Zackaria Chacko, Daniel Chung, Michael Graesser,
Salman Habib, Nemanja Kaloper, Hideo Kodama, Ann Nelson,
Maxim Perelstein,  Fernando Quevedo, Riccardo Rattazzi,
Yuri Shirman, Raman Sundrum, Takahiro Tanaka,
John Terning and Mark Wise for useful discussions.
We also thank Daniel Chung, Nemanja Kaloper, Maxim Perelstein and John 
Terning for helpful comments on the manuscript.
We would like to thank the Aspen Center for Physics where this work
was initiated during the workshop ``New Physics at the Weak Scale
and Beyond.''
C.C. is a J. Robert Oppenheimer fellow
at the Los Alamos National Laboratory. C.C., J.E.  are
supported by the US Department of Energy under contract W-7405-ENG-36.
C.G. is supported in part by the US Department of Energy under contract
DE-AC03-76SF00098 and in part by the National Science Foundation under
grant PHY-95-14797.


\end{document}